\documentclass[letterpaper, 10 pt, conference]{ieeeconf}  

\IEEEoverridecommandlockouts                              

\usepackage{graphics} 
\usepackage{epsfig} 
\usepackage{amsmath} 
\usepackage{lipsum}  
\usepackage{amssymb}  

\usepackage{algorithm2e} 
\RestyleAlgo{ruled}

\usepackage{cite}

\usepackage{xcolor}
\usepackage{listings} 
\definecolor{mGreenDTU}{rgb}{0,0.53,0.2}
\definecolor{mGray}{rgb}{0.5,0.5,0.5}
\definecolor{mPurple}{rgb}{0.58,0,0.82}
\definecolor{mRedDTU}{rgb}{0.6,0,0.0}
\definecolor{mBlueDTU}{rgb}{0.18, 0.24 ,0.95}
\definecolor{backgroundColour}{rgb}{0.95,0.95,0.95}
\lstdefinestyle{CStyle}{
    backgroundcolor=\color{backgroundColour},   
    commentstyle=\color{mGreenDTU},
    keywordstyle=\color{mRedDTU},
    numberstyle=\tiny\color{mGray},
    stringstyle=\color{mBlueDTU},
    basicstyle=\tiny,
    breakatwhitespace=false,         
    breaklines=true,                 
    captionpos=b,                    
    keepspaces=true,                 
    numbers=left,                    
    numbersep=4pt,                  
    showspaces=false,                
    showstringspaces=false,
    showtabs=false,                  
    tabsize=2,
    language=C,
    frame = single
}

\lstdefinestyle{CStyleInline}{
    backgroundcolor=\color{backgroundColour},   
    commentstyle=\color{mGreenDTU},
    keywordstyle=\color{mRedDTU},
    numberstyle=\tiny\color{mGray},
    stringstyle=\color{mBlueDTU},
    basicstyle=\normalsize,
    breakatwhitespace=false,         
    breaklines=true,                 
    captionpos=b,                    
    keepspaces=true,                 
    numbers=left,                    
    numbersep=5pt,                  
    showspaces=false,                
    showstringspaces=false,
    showtabs=false,                  
    tabsize=2,
    language=C,
    frame = single
}

\lstdefinestyle{PythonStyle}{
    backgroundcolor=\color{backgroundColour},   
    commentstyle=\color{mGreenDTU},
    keywordstyle=\color{mRedDTU},
    numberstyle=\tiny\color{mGray},
    stringstyle=\color{mBlueDTU},
    basicstyle=\normalsize,
    breakatwhitespace=false,         
    breaklines=true,                 
    captionpos=b,                    
    keepspaces=true,                 
    numbers=left,                    
    numbersep=5pt,                  
    showspaces=false,                
    showstringspaces=false,
    showtabs=false,                  
    tabsize=2,
    language=Python
}

\title{\LARGE \bf
Software principles and concepts applied in the implementation of cyber-physical systems for real-time advanced process control 
}

\author{Anders H. D. Andersen, Zhanhao Zhang, Steen Hørsholt, Tobias K. S. Ritschel, John Bagterp Jørgensen
\thanks{A. H. D. Andersen, Z. Zhanhao, S. Hørsholt, T. K. S Ritschel, and J. B. Jørgensen are with the Department of Applied
Mathematics and Computer Science, Technical University of Denmark, DK-2800 Kgs. Lyngby, Denmark.
Corresponding author: J. B. Jørgensen (E-mail: jbjo@dtu.dk).
        {\tt\small jbjo@dtu.dk}}%
}

\begin{document}

\maketitle
\thispagestyle{empty}
\pagestyle{empty}


\begin{abstract}

Cyber-physical systems (CPSs) for real-time advanced process control (RT-APC) are a class of control systems using network communication to control industrial processes. In this paper, we use simple examples to describe the software principles and concepts used in the implementation of such systems. The key software principles are 1) shared data in the form of a database, files, or shared memory, 2) timers and threads for concurrent periodic execution of tasks, and 3) network communication between the control system and the process, and communication between the control system and the internet, e.g., the cloud to enable remote monitoring and commands.
We show how to implement such systems for Linux operating systems applying the C programming language and we also comment on the implementation using the Python programming language.
Finally, we present a complete simulation experiment using a real-time simulator.

\end{abstract}


\section{Introduction}
\label{sec:Introduction}

Real-time advanced process control (RT-APC) enables the automatic, reliable, and efficient operation of complex industrial process systems. RT-APC relates to digital control methods such as proportional-integral-derivative (PID) control and model predictive control (MPC) executed periodically with real-time interval timers  \cite{Aastrom:Wittenmark:1997}.
Cyber-physical systems (CPSs) combine automatic control with network communication principles for remote control and monitoring of distributed processes.
A CPS for RT-APC (CPS-RT-APC) consists of numerous modules, e.g., controllers, user interfaces, and network communication modules that exchange data periodically and concurrently \cite{Jbar:etal:2018, Mois:etal:2015}. Consequently, implementations of such systems require real-time interval timers and threading principles as well as shared data and network communication concepts \cite{Gabier:2004, Wittenmark:Aastrom:Arzen:2022}.
Additionally, interval timers and threading principles enable the construction of real-time simulators as proxies for the physical plants. Therefore, these software principles enable real-time closed-loop simulations.

Software tools for simulations and development of real-time networked control systems exist: \cite{Ceven:etal:2003} applies \textit{TrueTime}, a MATLAB/Simulink-based simulation toolbox for real-time networked control systems, to a double tank system and \cite{Kim:Kumar:2010} describes real-time enhancements to \textit{Etherware}, a framework for
networked control systems, and applies this framework to an experimental setup of an inverted pendulum. Additionally, \cite{Bartusiak:etal:2022} describes an industry initiative called \textit{Open Process Automation} that defines a standards-based, open, secure, interoperable process automation architecture with an objective to provide the computing platforms for the improvement of industrial use of MPC and machine learning based process technologies.
However, the multidisciplinary nature of a CPS-RT-APC still makes it a challenging task to implement despite the existence of these software tools and standards, and the descriptions of the key principles of such a system in \cite{Aastrom:Wittenmark:1997, Gabier:2004, Wittenmark:Aastrom:Arzen:2022}.

In this paper, we demonstrate the key software principles and components in the implementation of a CPS-RT-APC. These software principles and components are 1) shared data, 2) timers and threads, and 3) network communication. We implement the shared data as databases, files, or shared memory. We apply the timers and threading principles for the concurrent and periodic execution of tasks. Finally, we apply network communication between the control system and the plant, and between the control system and the internet for cloud computing and remote monitoring. The novelty of our paper is the explicit demonstration of these software principles and components through code examples. We apply the C programming language in Linux operating systems (OS) and we also comment on the implementation using the Python high-level programming language. Finally, we present a simulation experiment for a CPS-RT-APC applied to a real-time simulator.

The remaining parts of this paper are structured as follows.
Section \ref{sec:SoftwareArchitecture} provides an overview of the CPS-RT-APC software architecture. Section \ref{sec:SharedData} presents the shared data for such a framework. Section \ref{sec:TimersAndThreading} describes the real-time interval timers and threading principles. In Section \ref{sec:Communication}, we describe the client-server architecture for network communication. Sections \ref{sec:ControlComputations} and \ref{sec:PlantAndRTsimulator} present implementations of a control module and a real-time simulator. Section \ref{sec:UserInterface} comments on the implementation of a user interface for the CPS-RT-APC. Section \ref{sec:RTSimulationExperiment} presents a simulation experiment for the CPS-RT-APC framework. Finally, Section \ref{sec:Conclusion} presents conclusions.


\section{Software Architecture}
\label{sec:SoftwareArchitecture}

\begin{figure}[tb]
    \centering
    \includegraphics[width=0.485\textwidth]{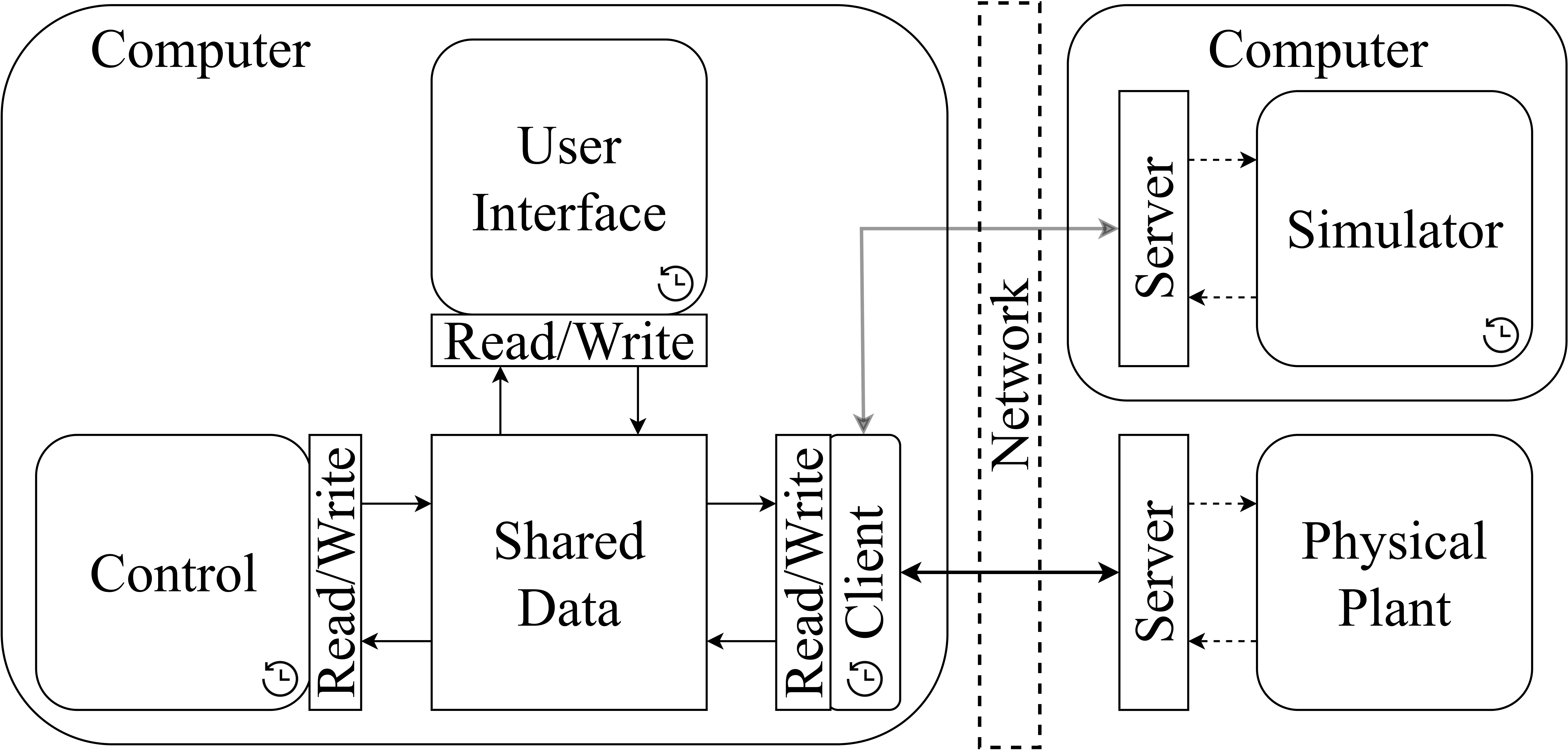}
    \caption{Schematic diagram of the CPS-RT-APC software architecture.}
    \label{fig:CPS_RT_APC}
\end{figure}

The CPS-RT-APC software architecture consists of modules that exchange data with shared data periodically. A module executes periodic tasks with a real-time interval timer and we execute the tasks of numerous modules concurrently by applying threading principles. The CPS-RT-APC framework applies a client-server network communication architecture for remote control and monitoring of physical plants. Such a client-server architecture also enables communication between the control systems and the internet, e.g., cloud computing. Additionally, interval timers, threading, and network communication principles enable the implementation of real-time simulators. We connect the servers to the physical plants and simulators, and we implement client modules with real-time interval timers for the periodic transmission of data. 
Fig. \ref{fig:CPS_RT_APC} presents a schematic diagram of the CPS-RT-APC software architecture with shared data and modules for a user interface, a controller, and a client. This client module connects to either a physical plant or a real-time simulator.

\section{Shared Data}
\label{sec:SharedData}

\begin{figure}[tb]
    \centering
    \includegraphics[width=0.485\textwidth]{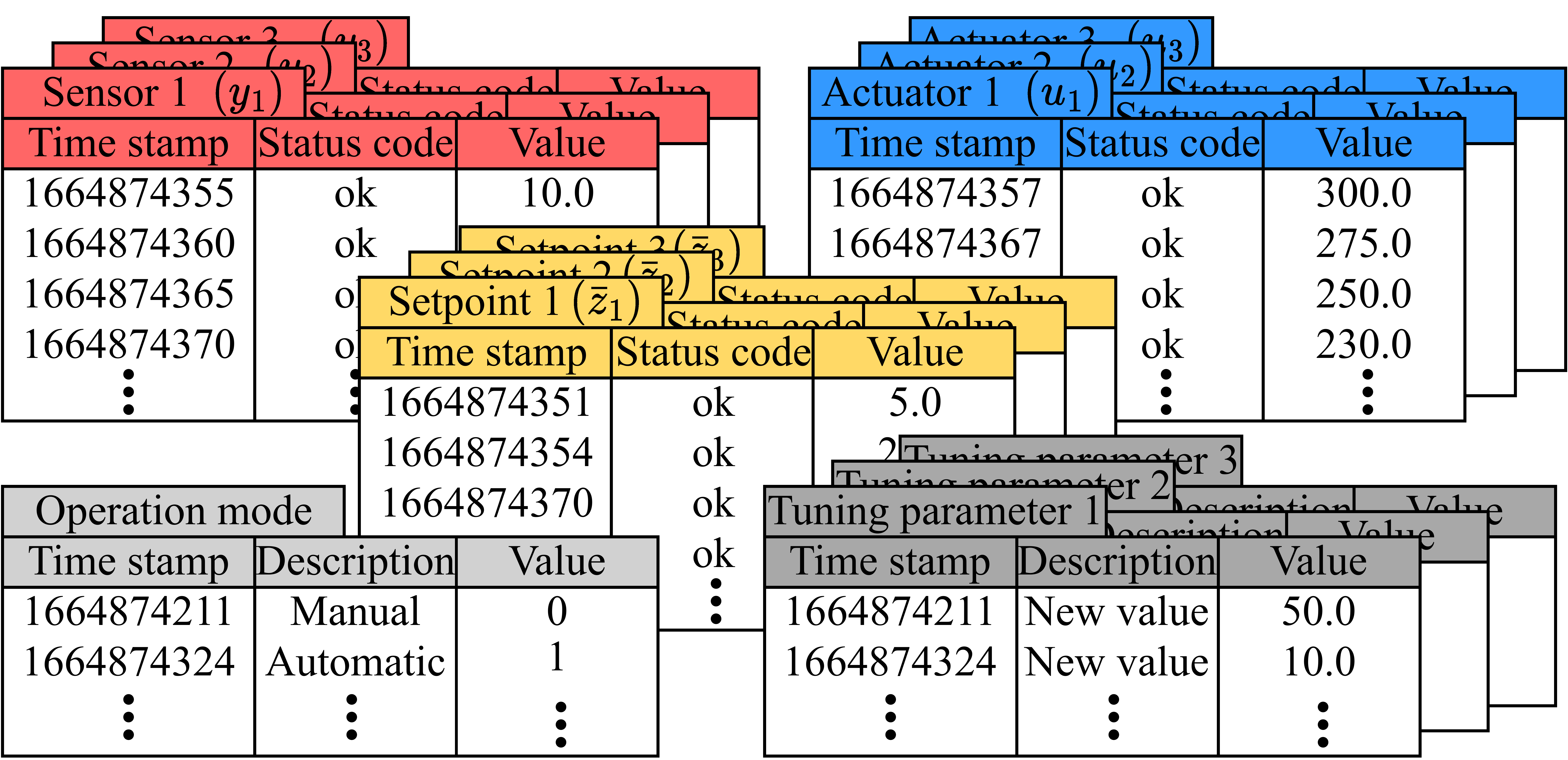}
    \caption{Illustration of the tables in shared data.}
    \label{fig:shared_data}
\end{figure}

We implement shared data as a collection of tables for sensor measurements, actuator commands, and setpoint values and we use $y$, $u$, and $\Bar{z}$ to represent these, respectively. We also represent control and operational configurations as tables. We may apply database systems, files, or shared memory to implement these tables. A single table represents the historical data of a unique variable and for each row in a table, we include a time stamp, a status code/description, and a numerical value.  We apply Unix time to represent the timestamp and we order the data in the tables in time. Fig.  \ref{fig:shared_data} illustrates tables for sensor and actuator data as well as setpoint and control configurations.
We apply mutual exclusion (mutex) principles to prevent a module from accessing a table in the shared data when another module is using it. 
We demonstrate the interaction with shared data using the PostgreSQL database system. Such a database system implements mutex-like principles. Consequently, we do not demonstrate mutex principles explicitly. For these demonstrations, we implement the source code \textit{apcshareddata} for wrapping \textit{libpg-dev}, a PostgreSQL C development library, into the functions insertIntoTableFloat() and readMultiRecentValsFloat() for writing and reading of tables, respectively. We use these functions in the real-time timer callback functions, as shown in the later sections.

\section{Timers and Threading}
\label{sec:TimersAndThreading}

\begin{figure}[tb]
    \centering
    \includegraphics[width=0.485\textwidth]{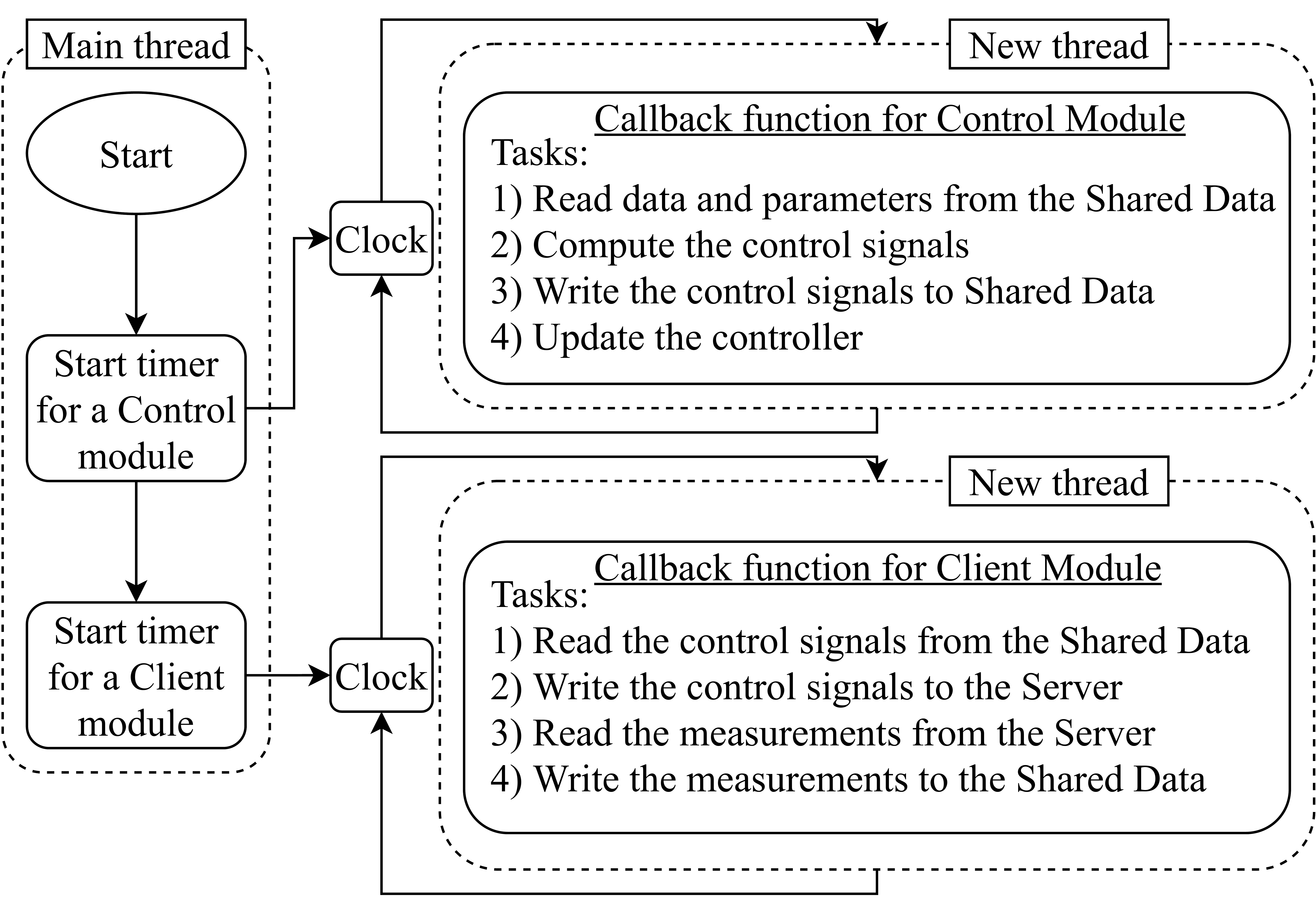}
    \caption{Flow diagram of tasks for control and client modules executed concurrently and periodically using interval timers with thread invocation.}
    \label{fig:timers_and_threads}
\end{figure}

Fig. \ref{fig:timers_and_threads} illustrates the concurrent and periodic execution of tasks for the control and client modules in Fig. \ref{fig:CPS_RT_APC} applying interval timers with thread invocation for the callback functions.
We may implement such interval timers with \textit{hard} real-time systems principles, e.g., real-time timers with high timing determinism and deadline guarantees. However, such systems rely on special-purpose software such as real-time operating systems (RTOS) \cite{Aastrom:Wittenmark:1997}. Alternatively, we may apply general-purpose operating systems (GPOS) such as the Linux OS for CPS-RT-APC systems that do not require such high timing determinism. We refer to such systems as \textit{soft} real-time systems and these will experience larger varying delays (jitter) compared to the hard real-time systems. 

We demonstrate the implementation of soft real-time system interval timers for the periodic and concurrent execution of user-defined tasks by applying timers from the portable operating system interface (POSIX) application programming interface (API), defined in \textit{POSIX.1b real-time extensions} for Linux OS. 
We configure the timers to invoke their callback functions in new threads to achieve concurrency of the timers \cite{Kerrisk:2010}.
We wrap these POSIX interval timers into functions for creating, starting, and stopping timers and we use structs to pass functions and function arguments to the callback functions through the \textit{sigevent} structure.
Listing \ref{lst:timers_example} presents an example for creating, starting, stopping, and deleting a POSIX interval timer with an interval of 2.5 s. Listing \ref{lst:timers_src} shows the \textit{apctimer} source code used to create, start, and stop POSIX timers.

\subsection{Python implementation}
\lstset{style=PythonStyle}
We use Python's Timer class to schedule the execution of tasks. This class enables the user to schedule the execution of these tasks in separate threads, similar to the POSIX interval timers configured with thread invocation of the callback functions.
 We create an instance of the Timer class as timer = Timer(interval, func, args=None, kwargs=None), where \textit{func} is the callback function and \textit{interval} is the start delay. We achieve periodic execution of the callback function by recursively applying the timer.start() and we apply timer.cancel() to stop the timers\cite{Hunt:2019}. Alternatively, we may design interval timers by applying the sleep function from the \textit{time} module together with loops. An implementation of such a method without accumulation of time drift is presented in \cite{Wittenmark:Aastrom:Arzen:2022}.
 \lstset{style=CStyleInline}



\begin{figure}[tb]
    \centering
\begin{minipage}{0.47\textwidth}
\begin{lstlisting}[caption={Example code for creating, starting, stopping, and deleting a timer.},label={lst:timers_example}, style=CStyle]
#include <stdio.h>
#include <stdlib.h>
#include <unistd.h>
#include <signal.h>      
#include "apctimer.h"

typedef struct Args_custom{
    void (*func)();     // Function to be called inside Callback function
    double interval;    // Interval of timer
}Args_custom;
void callback(union sigval sv){
    // Dereference sigval pointer
    Args_custom* args = (Args_custom*)sv.sival_ptr;
    // Call the function
    args->func(args->interval);
}
void printInterval(double interval){
    // Print interval of timer
    printf("Interval time for this timer is %0.2f s\n", interval);
}
int main(int argc, char const* argv[]){
    // Real-time timer interval
    double interval = 2.5; // seconds
    Args_custom *timer_args = malloc(sizeof(Args_custom));
    timer_args->func        = printInterval;
    timer_args->interval    = interval;
    // Create a timer that applies "callback" as the callback function
    timer_t* timer = create_apctimer(callback, timer_args);
    // Start timer (args: timerid, period)
    start_apctimer(timer, interval);
    // Sleep for 1 min
    sleep(60);
    // Stop the timer
    stop_apctimer(timer);
    // Delete the timer
    sleep(3);
    timer_delete(timer); // From POSIX.1b timer API
    // Free Args_custom argument
    free(timer_args);
    return 0;
}
\end{lstlisting}
\end{minipage}\hfill\\
\end{figure}

\begin{figure}[tb]
    \centering
\begin{minipage}{0.47\textwidth}
\begin{lstlisting}[caption={Source code for \textit{apctimer} applied in Listing \ref{lst:timers_example}.},label={lst:timers_src}, style=CStyle]
timer_t create_apctimer(void (*fnc)(), void* args){
    timer_t timerid;     // Timer identifier for POSIX.1b interval timer                               
    struct sigevent sev; // Structure for notification from async. routines                           
    memset(&sev, 0, sizeof(struct sigevent)); // Set struct members to 0
    sev.sigev_notify = SIGEV_THREAD;          // Config: Invoke callback function in new thread       
    sev.sigev_notify_function = fnc;          // Callback function       
    sev.sigev_value.sival_ptr = args;         // Arguments 
    // Create a real-time timer  using System-wide realtime clock    
    if(timer_create(CLOCK_REALTIME, &sev, &timerid) == -1){
        perror("timer_create");
    };                                                
    return timerid;
}
int start_apctimer(timer_t timerid, double Ts){
    int sec2nsec = 1000000000; // Convert seconds to nanoseconds 
    struct itimerspec ts;          
    int intpart = (int)Ts;         // Integer part of interval          
    double decpart = Ts - intpart; // Decimal part of interval
    memset(&ts, 0, sizeof(struct itimerspec)); // Set struct members to 0
    // Seconds and nanoseconds for timer interval 
    ts.it_interval.tv_sec  = intpart;                  
    ts.it_interval.tv_nsec = (int)(decpart*sec2nsec);
    // Start time for timer
    ts.it_value.tv_nsec = 1;    // Start time in nanoseconds. If = 0 then timer is disabled
    if(timer_settime(timerid, 0, &ts, NULL) == -1){
        perror("timer_settime");
        return 1;
    };
    return 0;
}
int stop_apctimer(timer_t timerid){
    struct itimerspec ts;
    // Set interval and start time to 0                          
    memset(&ts, 0, sizeof(struct itimerspec));  // Set struct members to 0    
    if(timer_settime(timerid, 0, &ts, NULL) == -1){ 
        perror("timer_settime");
        return 1;
    };
    return 0;
}
\end{lstlisting}
\end{minipage}\hfill\\
\end{figure}

\section{Network Communication}
\label{sec:Communication}
Industrial control systems apply network communication to transmit periodic and event-based signals such as control signals, measurements, and alarms between field devices, digital controllers, and operator interfaces. Network communication standards such as the open platform communications unified architecture (OPC UA) standardize the communication between industrial components and many industries apply this standard.
The OPC UA standard applies a client-server communication architecture and offers a variety of transport protocols such as TCP/IP based protocols for reliable data transmission \cite{Galloway:Hancke:2013, Mois:etal:2015}.

In this paper, we limit the discussion to the periodic transmission of data between the plant and control system by applying a client-server communication architecture with a TCP/IP protocol. We present such a communication architecture for the CPS-RT-APC framework in C with the internet domain stream sockets applied to a real-time simulator. We apply the functions packMultiMsg() and unpackMultiMsg() defined in the source code \textit{apcsockets} to pack and unpack data from numerous processes into single strings and we transmit these strings between the client and the server. We also apply this library for the initialization of the clients and servers. We do not show the source code but \cite{Kerrisk:2010} provide code examples for the initialization of these internet domain stream socket clients and servers and how communication between them can be made. We apply real-time interval timers for the client module for the periodic transmission of this data and we pass a pointer to a shared struct between the server and the real-time simulator. 

\subsection{Client}
Listing \ref{lst:client_timer_code} demonstrates how to initialize and connect a client to a server. We demonstrate how to apply an interval timer for the periodic transmission of information between the shared data and the server. The timer callback function is client() and it requires connections to the shared data and to the server. We pass these as arguments to the callback function by using the struct \lstinline{Args_client}. 
We choose the timer interval of a client module to be smaller compared to the timer interval of a control module. We do this to ensure a fast update of the data between the plant and the control module.

\begin{figure}[htb]
    \centering
\begin{minipage}{.47\textwidth}
\begin{lstlisting}[caption={Example code for client module that connects to a Server at Local-host with port number 43051. The main function does not show how to stop and delete the interval timer for the Client module.},label={lst:client_timer_code}, style=CStyle]
#include <stdio.h>
#include <stdlib.h>
#include <string.h>
#include <libpq-fe.h>
#include <signal.h>
#include <arpa/inet.h>      
#include <sys/socket.h>
#include "apctimer.h"
#include "apcshareddata.h" 
#include "apcsockets.h"

typedef struct Args_client{
    int sock_conn;      // Connection to socket Server
    PGconn* db_conn;    // Connection to PostgreSQL database (Shared Data)
}Args_client;
void client(union sigval sv){
    // Init buffers for sending and receiving messages
    char buff_send[1024] = {0}, buff_recv[1024] = {0}; 
    // Dereference pointer to timer data struct
    Args_client* args = (Args_client*)sv.sival_ptr;
    // Dimensions {1) measurements, 2) setpoints, 3) manipulated variables}
    int n[3] = {1,1,1};
    // Read number of inputs and outputs from database
    readMultiRecentValsInt(args->db_conn, "dim", 3, NULL, NULL, n);
    // Arrays for storing numerical values, timestamps, status codes
    double y[n[0]], u[n[2]];    
    char ts_send[n[2]][1024], stat_send[n[2]][1024]; // timestamp and status
    char ts_recv[n[0]][1024], stat_recv[n[0]][1024]; // timestamp and status
    // Read actuator values from the Shared Data
    readMultiRecentValsFloat(args->db_conn, "actuator", n[2], ts_send, stat_send, u);
    // Pack actuator data as a string.
    memset(buff_send, 0, sizeof(buff_send));
    packMultiMsg(buff_send, n[2], ts_send, stat_send, u);
    // Send string message to the Server
    send(args->sock_conn, buff_send, strlen(buff_send), MSG_NOSIGNAL); 
    // Init and read string message from the Server
    memset(buff_recv, 0, sizeof(buff_recv));
    read(args->sock_conn, buff_recv, 1024);  
    // Unpack sensor data from the Server into arrays
    unpackMultiMsg(ts_recv, stat_recv, y, buff_recv);
    // Write sensor data from the Server to the Shared Data
    for(int i = 0; i < n[0]; i++)
        insertIntoTableFloat(args->db_conn, "sensor", i+1, ts_recv[i], stat_recv[i], y[i]); 
}
int main(int argc, char const* argv[]){
    // Get the ip-address and port number of the Server
    char serv_ip[128] = "127.0.0.1"; // Local-host
    int portnumber = 43051; // Example of Port number
    // Real-time timer interval
    double interval = 1.0; // seconds
    // Create connection to PostgreSQL database using conn_str.
    PGconn *db_conn = PQconnectdb(conn_str);
    // Init tables in database using db_conn
    // Create file descriptor for client
    int client_fd = client_init();
    // Connect to the Server (args: IP address, portnumber)
    client_connect(client_fd, serv_ip, portnumber);
    // Create Args_client-struct for the Client callback
    Args_client *args = malloc(sizeof(Args_client));
    args->sock_conn = client_fd;
    args->db_conn   = db_conn;
    // Create a timer for the Client module
    timer_t* clientTimer = create_apctimer(client, args);
    // Start timer (args: timerid, period)
    start_apctimer(timer, interval);
    // Stop and delete timer and free "args" (NOT SHOWN)
    // Close socket
    close(client_fd);
    return 0;
}

\end{lstlisting}
\end{minipage}\hfill\\
\end{figure}

\subsection{Server}

We show the implementation of a server that reads the sensor measurements and write actuator commands from and to a real-time simulator. We implement such a server as an infinite loop that waits for requests from a client. Listing \ref{lst:server_code_main} demonstrates a server for the CPS-RT-APC with server() representing the infinite loop. We pass the data for process variables applying the custom struct Args\_sim\_plant. We use pointers to pass the data of such a struct to both the server and real-time simulator. Such an implementation requires mutex-like principles to synchronize the communication between the server and the real-time simulator and we demonstrate this by locking and unlocking the code where server() manipulates the data in the shared struct.
\begin{figure}[tb]
    \centering
\begin{minipage}{.46\textwidth}
\begin{lstlisting}[caption={Example code for a server connected to a real-time simulator.  The server and real-time simulator apply the Args\_sim\_plant-struct as shared memory for measurements and manipulated variables.},label={lst:server_code_main}, style=CStyle]
#include <stdio.h>
#include <stdlib.h>
#include <string.h>
#include <libpq-fe.h>
#include <sys/time.h>
#include <pthread.h>
#include <netinet/in.h>     
#include <sys/socket.h>     
#include "apcshareddata.h" 
#include "apcsockets.h"

// Mutex for writing/reading of measurements
pthread_mutex_t mutex; 

void server(Args_sim_plant* args, int sock_conn){
    // Init buffers for sending and receiving messages
    char buff_recv[1024] = {0}, buff_send[1024] = {0};        
    char ts[27]; // timestamp
    // Arrays for storing numerical values, timestamps, and status codes
    double y[args->n[1]], u[args->n[2]];    
    char ts_send[args->n[1]][1024], stat_send[args->n[1]][1024]; 
    char ts_recv[args->n[2]][1024], stat_recv[args->n[2]][1024];
    while(1){
        // Read string message from the Client
        if (read(sock_conn, buff_recv, 1024) < 0){
            printf("Client disconnected...\n");
            break;
        };
        // Insert numerical values for actuator from the Client into arrays          
        unpackMultiMsg(ts_recv, stat_recv, u, buff_recv);
        for(int i = 0; i < args->n[2]; i++)
            args->u[i] = u[i];
        // Lock shared memory
        pthread_mutex_lock(&mutex);
        // Get measurements with timestamps and status
        for(int j = 0; j < args->n[1]; j++){
            genTimeStampUTC(ts);
            y[j] = args->y[j];
            strcpy(ts_send[j], ts); 
            strcpy(stat_send[j], "'ok'"); // Write status code
        }
        // Unlock shared memory
        pthread_mutex_unlock(&mutex);
        // Pack sensor data as string.
        memset(buff_send, 0, sizeof(buff_send));
        packMultiMsg(buff_send, args->n[1], ts_send, stat_send, y);
        // Send string message to the Client
        if (send(sock_conn, buff_send, strlen(buff_send), MSG_NOSIGNAL) < 0){                            
            printf("Client disconnected...\n");
            break;           
        }                                           
    }
}
int main(int argc, char const* argv[]){
    // Create a custom struct for the Server and the Simulator callback
    Args_sim_plant *argsSimulator = malloc(sizeof(Args_sim_plant));
    // Fill in members of Args_sim_plant-struct (NOT SHOWN)
    // Init server
    int server_fd, new_fd;
    // Length of queue for clients (Number chosen arbitrarily)
    int client_queue = 3;
    server_fd = server_init();
    // Listening for clients
    server_listen(server_fd, client_queue);
    // Accept client communication with the Server
    new_fd = server_accept(server_fd);
    // Start the Server
    server(argsSimulator, new_fd);
    // Close socket
    close(new_fd);
    // Free argsSimulator (NOT SHOWN)
    return 0;
}
\end{lstlisting}
\end{minipage}\hfill\\
\end{figure}

\begin{figure}[tb]
    \centering
    \includegraphics[width=0.485\textwidth]{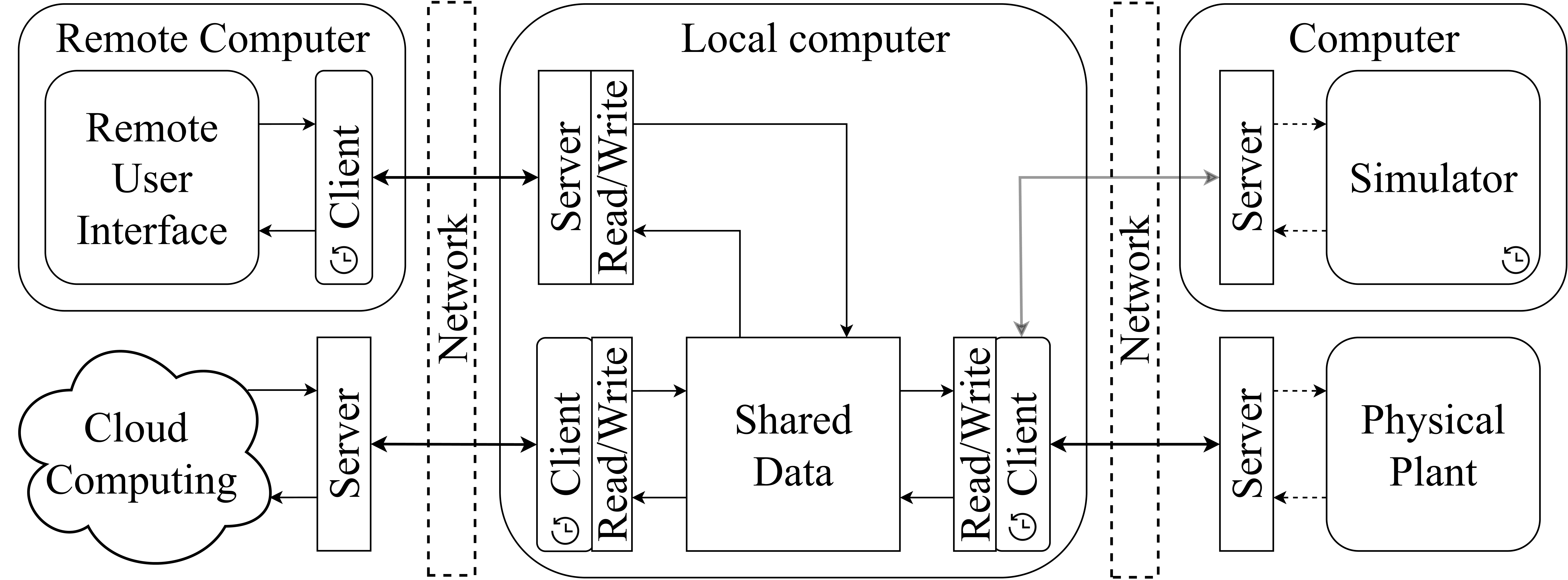}
    \caption{Schematic diagram of the CPS-RT-APC software architecture with remote monitoring and cloud computing.}
    \label{fig:CPS_RT_APC2}
\end{figure}

Finally, such a client-server communication architecture enables cloud computing, storage, and remote monitoring of CPS-RT-APC systems. Fig. \ref{fig:CPS_RT_APC2} illustrates how a client-server architecture for the CPS-RT-APC may enable the usage of remote computers for monitoring and cloud computing.

\subsection{Python Implementation}
\lstset{style=PythonStyle}
Python's socket communication mechanism is similar to the socket functions applied in the code examples in Listings \ref{lst:client_timer_code} and \ref{lst:server_code_main}, i.e.,  sock = socket.socket(socket.AF\_INET, socket.SOCK\_STREAM) creates a socket server. 
We apply sock.bind(server\_address) to bind the socket to the host-machine and we use  sock.listen() to listen for clients. To connect a client to a server, we create a client socket with \lstinline{sock = socket.socket()} and then apply sock.connect(ipaddress, port) \cite{Hunt:2019}.
\lstset{style=CStyleInline}



\section{Control computations}
\label{sec:ControlComputations}
Fig. \ref{fig:timers_and_threads} outlines the fundamental tasks in a control module for the CPS-RT-APC and we demonstrate the implementation of these tasks in this section. However, it is also common to see logic tasks in the form of evaluating the state of actuators and sensors included in such a module. Listing \ref{lst:timer_code_control} presents an example of such an implementation applying the callback function control(). This callback function requires the control function, the controller update function, the sampling time, and a connection to the shared data. We pass these as arguments to the callback function by using the struct Args\_control.

\begin{figure}[tb]
    \centering
\begin{minipage}{0.47\textwidth}
\begin{lstlisting}[caption={Example code for the implementation of a callback function for a control module.},label={lst:timer_code_control}, style=CStyle]
#include <stdio.h>
#include <libpq-fe.h>
#include "apctimer.h" 
#include "apcshareddata.h" 

typedef struct Args_control{
    void (*cfunc)();        // controller function
    void (*cfunc_update)(); // controller update function 
    double Ts;              // Sampling time
    PGconn* db_conn;        // Connection to PostgreSQL database
}Args_control;
void cfunc(double *u, double *y, double* z_bar, PGconn* db_conn){
    /* Unpack varargin for setpoints 
    Compute control signal
    u[0] = ...
    u[1] = ... */
}
void cfunc_update(double *u, double *y, double* z_bar, PGconn* db_conn){
    /* Update controller */
}
void control(union sigval sv){
    int op_mode = 0;
    char ts[27]; // timestamp
    // Dereference sigval pointer
    Args_control* args = (Args_control*)sv.sival_ptr;
    // Get operation mode
    readRecentValInt(args->db_conn, "opmode", 0, NULL, NULL, &op_mode);
    // Update control algorithm operation mode = 1 (automatic)
    if(op_mode){     
        // Read setpoint and measurements from Shared Data
        // Dimensions: 1) measurements, 2) setpoints, 3) manipulated variables
        int n[3] = {1,1,1}; 
        readMultiRecentValsInt(args->db_conn, "dim", 3, NULL, NULL, n);
        // Measurements, setpoints, and manipulated variables
        double y[n[0]], z_bar[n[1]], u[n[2]];
        // Read only numerical values of sensor and setpoint
        readMultiRecentValsFloat(args->db_conn, "sensor", n[0], NULL, NULL, y);
        readMultiRecentValsFloat(args->db_conn, "setpoint", n[1], NULL, NULL, z_bar);
        // Control algorithm
        args->cfunc(u, y, z_bar, args->db_conn);
        genTimeStampUTC(ts); // Get current Unix time
        // Write manipulated variables to Shared Data
        for(int i = 0; i < n[2]; i++)
            insertIntoTableFloat(args->db_conn, "actuator",i+1, ts, "'ok'", u[i]);
        // Update controller
        args->cfunc_update(u, y, z_bar, args->db_conn);
    }
}
int main(int argc, char const* argv[]){
    // Interval of timer for Control module
    double Ts = 2.0; // seconds
    // Create connection to PostgreSQL database
    char conn_str[]  = "user=apc_user dbname=apc_db password=apc_password";
    PGconn *db_conn = PQconnectdb(conn_str);
    // Init tables in database using db_conn 
    // Create struct for controller callback function
    Args_control *args = malloc(sizeof(Args_control));
    // Fill in members of Args_control-struct 
    // Create a timer for the Controller
    timer_t* cntrTimer = create_apctimer(control, args);
    // Start the timer. Stop and delete the timer when finish (NOT SHOWN);
    free(args);
    // Close connection to db
    PQfinish(db_conn);
    // Free Args_control argument
    free(args);
    return 0;
}
\end{lstlisting}
\end{minipage}\hfill\\
\end{figure}
\section{Real-time Simulator}
\label{sec:PlantAndRTsimulator}

In this section, we demonstrate the implementation of a real-time simulator applying a POSIX interval timer.
Listing \ref{lst:simulator} presents an example of such an implementation applying the callback function simulator(). This callback function requires a mathematical model of the plant, the states, inputs, and output dimensions, model parameters and control signals, and an ordinary differential equation (ODE) solver. We pass these as arguments to the callback function through the struct Args\_sim\_plant.

\begin{figure}[tb]
    \centering
\begin{minipage}{.47\textwidth}
\begin{lstlisting}[caption={Example code for the implementation of a callback function for a real-time simulator. The Main function code is not presented.},label={lst:simulator}, style=CStyle]
#include <stdio.h>
#include <stdlib.h>
#include <signal.h>  
#include <libpq-fe.h>
#include "apctimer.h"
#include "apcshareddata.h" 

struct varargin{
    void *args; // Vector of arguments
    int nargs;  // Number of arguments
};
typedef struct Args_sim_plant{
    void (*ffunc)();            // x' = f(t, x, args)
    void (*gfunc)();            // 0  = g(t, x, args)
    struct varargin arg_vec;    // argument vector for f and g
    double Ts;                  // sampling time
    double *x;                  // states
    double *y;                  // measurements
    double *u;                  // manipulated variables
    int *n;                     // dimensions: (states, outputs, inputs)
    void (*odesolver)();        // ODE solver function
    int N;                      // Number of steps in ODE solver
}Args_sim_plant;

void ffunc(double t, double* x, double* dx, double* varargin){
    /* Unpack parameters and manipulated variables from varargin */
    /* Compute dx/dt
    dx[0]   = ....
    dx[1]   = .... */
}
void gfunc(double* x, double* y, double* varargin){
    /* Unpack parameters and manipulated variables from varargin */
    /* Compute y
    y[0]   = ....
    y[1]   = .... */
}
void odesolver( void (*ffunc)(), double t0,  double tN, int N,
                double* x0,int nx,double* varargin,
                double (*X)[N+1],  double T[N]){
    // Compute the solution T, X
}
void simulator(union sigval sv){
    // Dereference sigval pointer
    Args_sim_plant* args = (Args_sim_plant*)sv.sival_ptr;
    // Number of states and manipulated variables
    int nx = args->n[0], nu = args->n[2]; 
    // ODE solver configurations
    int N = args->N; // Steps for ODE solver
    double X[nx][N+1], T[N+1];      // Arrays for ode solution
    double t0 = 0.0, tf = args->Ts; // Integration time
    double *arg = args->arg_vec.args;
    int nargs   = args->arg_vec.nargs;
    // Init vector for parameters and manipulated variables
    double varargin[nargs+nu];
    // Parameters  
    for(int i = 0; i < nargs; i++)
        varargin[i] = arg[i];
    // Manipulated variables
    for(int k = 0; k < nu; k++)
        varargin[nargs+k] = args->u[k];
    // Solve ODE from timspan [t0 tf] with N steps
    args->odesolver(args->ffunc, t0, tf, N, args->x, nx, varargin, X, T);
    // Update the states
    for(int j = 0; j < nx; j++)    
        args->x[j] = X[j][N];    
    // Compute measurements
    args->gfunc(args->x, args->y, varargin);
}
int main(int argc, char const* argv[]){
    // Init parameters of model
    // Create Args_sim_plant-struct for the Simulator callback using
    // Fill in members of Args_sim_plant-struct
    // Create, start, stop, and delete af timer with simulator() as callback
    return 0;
}
\end{lstlisting}
\end{minipage}\hfill\\

\end{figure}
\section{User Interface}
\label{sec:UserInterface}

\begin{figure}[tb]
    \centering
    \includegraphics[width=0.485\textwidth]{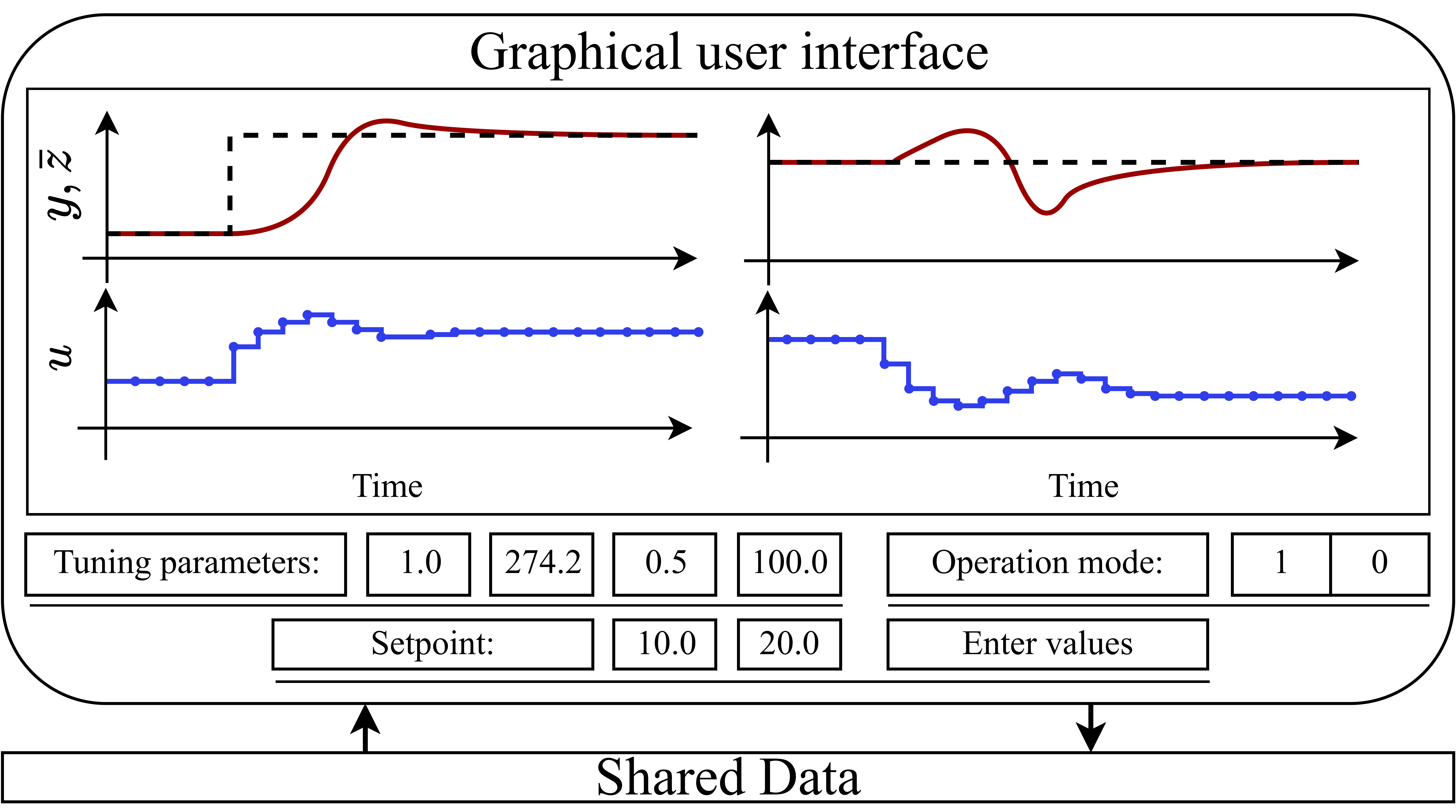}
    \caption{Illustration of a graphical user interface for the CPS-RT-APC.}
    \label{fig:user_interface}
\end{figure}

A user interface module enables an operator to view process data through real-time plotting of measurements and manipulated variables, and to influence the control system through manipulation of setpoints and control configurations.
Such a user interface requires a connection to the shared data for exchanging information between the other modules as well as real-time interval timers for real-time plotting and an event-based mechanism for changing setpoints. Fig. \ref{fig:user_interface} illustrates a graphical user interface for a CPS-RT-APC with real-time plotting of data from two sensors and two actuators and buttons and text-fields for manipulating setpoints, tuning parameters, and the operation mode.

\section{Real-time simulation experiments}
\label{sec:RTSimulationExperiment}

\begin{figure}[tb]
    \centering
    \includegraphics[width=0.49\textwidth]{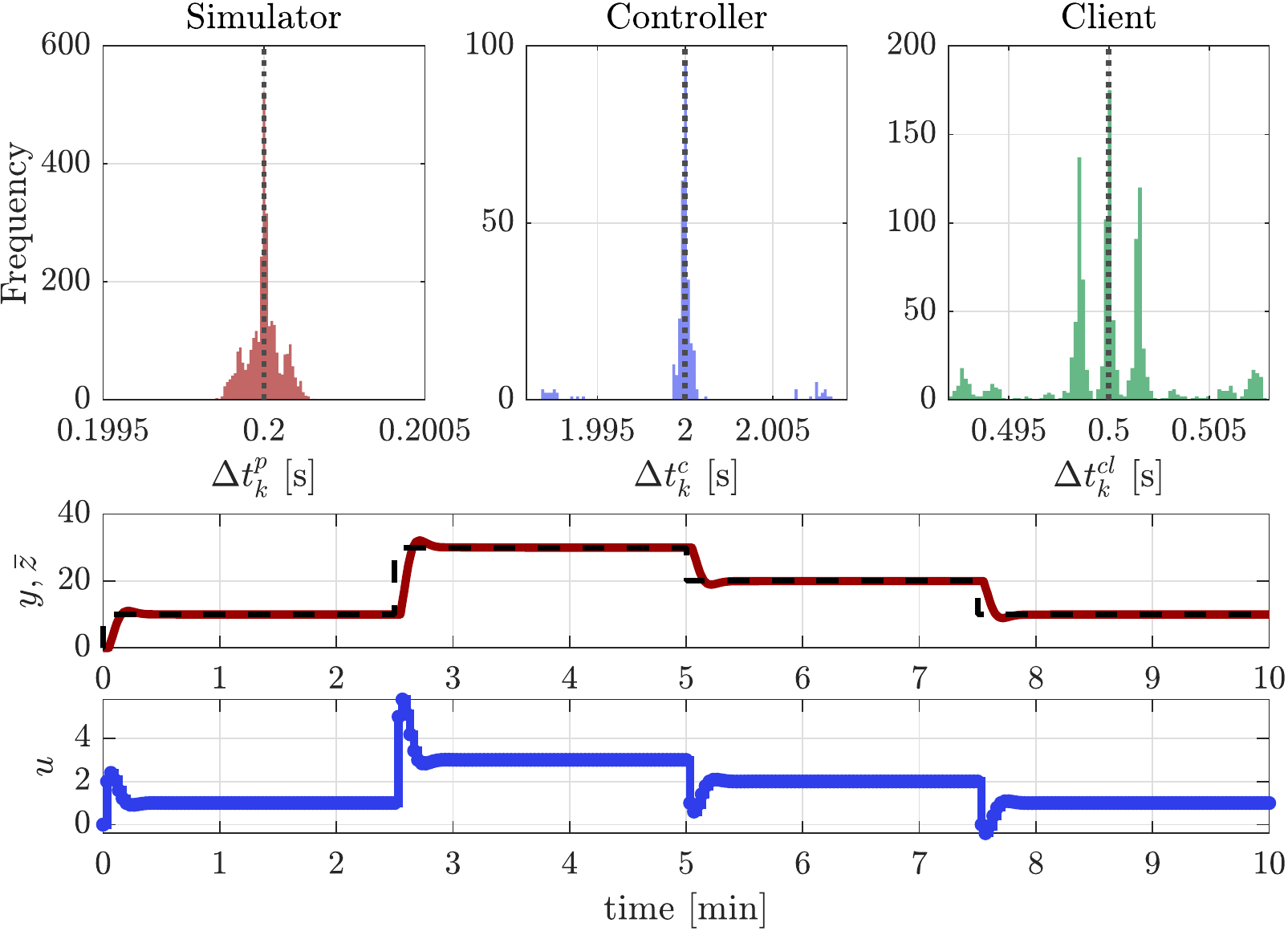}
    \caption{Top) histograms of jitter for the simulator, control module, and client module with mean time increments.  Bottom) the setpoints, measurements, and manipulated variables for the simulation experiment.}
    \label{fig:example_combined_plot}
\end{figure}

We test the CPS-RT-APC framework with a real-time simulator of the ODE $\dot x(t) = (1/\tau)(-x(t)+Ku(t))$ and with measurement equations $y(t_k) = x(t_k)$ and output equation $z(t) = x(t)$, where $t$ is time, $x(t)$ is the state, $u(t)$ is the manipulated variable, $y(t_k)$ is the measurement, and $z(t)$ is the controlled variable. We choose the parameters $K = 10.0$ and $\tau = 10.0$ and the simulator timer interval $T_s^{p}$ = 0.2 s. We apply a zero-order hold parameterization of the manipulated variable $u(t) = u_k$ for $t_{k}^{p} \leq t \leq t_{k+1}^{p}$ with $t_{k+1}^{p} = t_{k}^{p}+T_s^{p}$ being the time increment for the simulator.
We track a time-varying setpoint, $\bar{z}_k$, with the PI-control strategy $e_k = \bar{z}_k-y_k$, $P_k = K_pe_k$, $u_k = \bar{u}_k + P_k + I_k$, $I_{k+1} = I_{k}+({K_pT_s^c}/{\tau_i})e_k$, and $I_0 = \bar{I}$ where $T_s^c$ is the interval time of the controller, and $K_p$, $\tau_i$, and $\bar{u}_k$ are the proportional gain, integral time constant, and chosen operating point for $u_k$, respectively. We choose the controller parameters $K_p = 0.2$, $\tau_i = 10.0$, $\bar{u} = 0.0$, and the interval time $T_s^c = 2.0$ s. We update a sequence of setpoints applying a real-time timer with interval $T_s^{\bar{z}} = 150.0$ s. We apply a client module with interval $T_s^{cl} = 0.5$ s and we start the timers simultaneously. We execute the control system and the simulator on the same host computer. Finally, we apply a PostgreSQL database system for storing the data for $y(t_k), u_k$ and $\bar{z}_{k}$. Fig. \ref{fig:example_combined_plot} presents histograms for the jitter $\Delta t_k^{i} = t_{k+1}^i-t_{k}^i$ where $t_{k+1}^i = t_k^i+T_s^i$ for $i\in\{p, c, cl\}$ as well as the process variables of the simulation experiment.

\section{Conclusion}
\label{sec:Conclusion}

In this paper, we demonstrate the key software principles and components applied in the implementation of cyber-physical systems for real-time advanced process control. These principles and components are shared data, timers and threads for concurrent periodic execution of tasks, and network
communication between the control system and the process and between the control system and the internet. We demonstrate these principles and components with code examples in the C programming language for Linux operating systems. We apply these examples to a real-time simulation experiment of an ODE, where a PI-control strategy tracks a time-varying setpoint. The results show high timing precision and good setpoint tracking.

\addtolength{\textheight}{-12cm}   





\bibliographystyle{IEEEtran}
\bibliography{ref/CPS_RT_APC.bib}

\begin{thebibliography}{10}
\providecommand{\url}[1]{#1}
\csname url@rmstyle\endcsname
\providecommand{\newblock}{\relax}
\providecommand{\bibinfo}[2]{#2}
\providecommand\BIBentrySTDinterwordspacing{\spaceskip=0pt\relax}
\providecommand\BIBentryALTinterwordstretchfactor{4}
\providecommand\BIBentryALTinterwordspacing{\spaceskip=\fontdimen2\font plus
\BIBentryALTinterwordstretchfactor\fontdimen3\font minus
  \fontdimen4\font\relax}
\providecommand\BIBforeignlanguage[2]{{%
\expandafter\ifx\csname l@#1\endcsname\relax
\typeout{** WARNING: IEEEtran.bst: No hyphenation pattern has been}%
\typeout{** loaded for the language `#1'. Using the pattern for}%
\typeout{** the default language instead.}%
\else
\language=\csname l@#1\endcsname
\fi
#2}}

\bibitem{Aastrom:Wittenmark:1997}
K.~J. Åström and B.~Wittenmark, \emph{Computer-Controlled Systems. Theory and
  Design}, 3rd~ed.\hskip 1em plus 0.5em minus 0.4em\relax Prentice-Hall, Inc.,
  1997.

\bibitem{Jbar:etal:2018}
M.~Jbair, B.~Ahmad, M.~H. Ahmad, and R.~Harrison, ``Industrial cyber physical
  systems: A survey for control-engineering tools,'' in \emph{1st IEEE
  International Conference on Industrial Cyber-Physical Systems, ICPS 2018},
  Saint Petersburg, Russia, May 2018, pp. 270--276.

\bibitem{Mois:etal:2015}
G.~Mois, S.~Folea, and T.~Sanislav, ``Communication in cyber-physical
  systems,'' in \emph{19th International Conference on System Theory, Control
  and Computing, ICSTCC 2015 - Joint Conference SINTES 19, SACCS 15, SIMSIS
  19}, Cheile Gradistei, Romania, Oct. 2015, pp. 303--307.

\bibitem{Gabier:2004}
A.~Gambier, ``Real-time control systems: a tutorial,'' in \emph{2004 5th Asian
  Control Conference}, Melbourne, Australia, July 2004, pp. 1024--1031.

\bibitem{Wittenmark:Aastrom:Arzen:2022}
B.~Wittenmark, K.-J. Åström, and K.-E. Årzén, \emph{Computer Control: An
  Overview}, ser. IFAC Professional Briefs.\hskip 1em plus 0.5em minus
  0.4em\relax IFAC - International Federation of Automatic Control, 2022.

\bibitem{Ceven:etal:2003}
A.~Cevin, D.~Henriksson, B.~Lincoln, J.~Eker, and K.-E. Årzén, ``How does
  control timing affect performance? {A}nalysis and simulation of timing using
  jitterbug and truetime,'' \emph{IEEE Control Systems Magazine}, vol.~23,
  no.~3, pp. 16--30, 2003.

\bibitem{Kim:Kumar:2010}
K.-D. Kim and P.~R. Kumar, ``Design and experimental verification of real-time
  mechanisms for middleware for networked control,'' in \emph{2010 American
  Control Conference}, Baltimore, MD, USA, June 2010, pp. 2119--2124.

\bibitem{Bartusiak:etal:2022}
R.~D. Bartusiak, S.~Bitar, D.~L. DeBari, B.~G. Houk, D.~Stevens,
  B.~Fitzpatrick, and P.~Sloan, ``Open process automation: A standards-based,
  open, secure, interoperable process control architecture,'' \emph{Control
  Engineering Practice}, vol. 121, p. 105034, 2022.

\bibitem{Kerrisk:2010}
M.~Kerrisk, \emph{The Linux programming interface: A Linux and UNIX System
  Programming Handbook}, 1st~ed.\hskip 1em plus 0.5em minus 0.4em\relax San
  Francisco, CA: No Starch Press, 2010.

\bibitem{Hunt:2019}
J.~Hunt, \emph{Advanced Guide to Python 3 Programming (Undergraduate Topics in
  Computer Science)}, 1st~ed.\hskip 1em plus 0.5em minus 0.4em\relax
  Gewerbestrasse 11, 6330 Cham, Switzerland: Springer Nature Switzerland, 2019.

\bibitem{Galloway:Hancke:2013}
B.~Galloway and G.~P. Hancke, ``Introduction to industrial control networks,''
  \emph{IEEE Communications Surveys \& Tutorials}, vol.~15, no.~2, pp.
  860--880, 2013.

\end{thebibliography}


\end{document}